\documentclass[pre,showpacs,reprint,nofootinbib]{revtex4-1}
\def\P{\partial}
\usepackage{amssymb}
\usepackage{amsmath}
\usepackage{graphicx}
\usepackage{color}

\newcommand{\be}{\begin{equation}}
\newcommand{\ee}{\end{equation}}

\newcommand{\py}{\partial_y}
\newcommand{\pz}{\partial_z}

\begin{document}

\title{A bilateral shear layer between two parallel Couette flows}
\author{Vagesh D. Narasimhamurthy}
\author{Simen \AA.\ Ellingsen} 
\author{Helge I. Andersson}
\affiliation{Fluids Engineering Division, Department of Energy and Process Engineering, 
Norwegian University of Science and Technology (NTNU), N-7491 Trondheim, Norway}


\begin{abstract}
We consider a shear layer of a kind not previously studied to our knowledge. Contrary to the classical free shear
layer, the width of the  shear zone does not vary  in the streamwise direction but rather exhibits a lateral variation. Based on some 
simplifying assumptions, an analytic solution has been derived for the new shear layer. These assumptions have been justified by a comparison with numerical solutions of the 
full Navier-Stokes equations, which accord with the analytical solution to better than $1\%$ in the entire domain. An explicit formula is found for the width of the shear zone as a function of wall-normal coordinate. This width is independent of 
wall velocities in the laminar regime. Preliminary results for a co-current laminar-turbulent   
shear layer in the same geometry are also presented. Shear-layer instabilities were then developed and resulted in an unsteady mixing zone at the interface between the two co-current streams.
\end{abstract}

\maketitle

\section{Introduction}

\begin{figure}[t]
\centering
\includegraphics[width=0.4\textwidth]{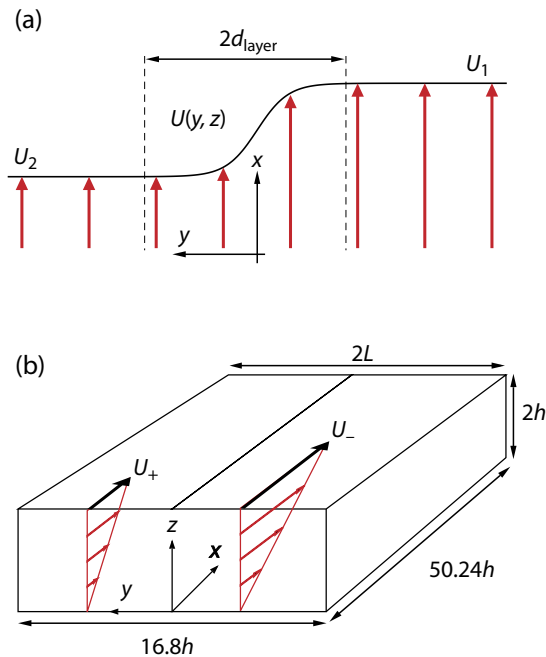}
\caption{(Color online) A shear layer between two parallel streams. (a) Generic sketch of a planar shear layer between streams with velocities $U_1$ and $U_2$; and (b) two co-current planar Couette flows with wall speeds $U_-$ and $U_+$. The dimensions of the flow configuration refer to those used in the numerical calculations.}
\label{domain}
\end{figure}

The laminar boundary layer which forms in the 
shear region between two semi-infinite 
uniform parallel streams admits a Blasius type of similarity solution. Lock \cite{Lock} 
considered streams of two different fluids and with different velocities, say $U_2$ and 
$U_1$, as depicted in figure \ref{domain}(a). His numerical solutions are reproduced 
in textbooks on viscous flows, e.g. Panton 
\cite{Panton} and White \cite{White}. In the special case where the two fluids are the same 
Lock \cite{Lock} found that the solution depended only on the ratio $U_2/U_1$ of the 
velocities of the two streams. A particular feature of the plane 
shear layer (also referred to as the classical mixing layer) is that the width of the shear
zone thickens as $(x \nu/U_1)^{1/2}$, where $\nu$ is the 
kinematic viscosity and $x$ is the streamwise distance from the trailing end of the plate 
where the two semi-infinite streams merge. Klemp and Acrivos \cite{Klemp} considered the 
non-uniqueness of this boundary layer problem due to an indeterminacy of the streamline 
separating the two streams. Even by including all higher-order effects in their analysis, 
the position of the dividing streamline remained indeterminate.

A special feature of the plane shear layer is that stability analyses, for instance that 
by Betchov and Szewczyk \cite{Betchov}, show that the flow is unstable at all Reynolds numbers. 
Bhattacharya et al. \cite{Bhattacharya} ascribed this peculiarity to the parallel-flow assumption 
on which the Orr-Sommerfeld equation is based. They therefore formulated a non-parallel stability 
problem and found a critical Reynolds number close to 30, below which the flow is convectively stable.

The stability of the flow in a circular shear zone was investigated by Rabaud and Couder \cite{Rabaud}. They developed an experimental apparatus in which the fluid was enclosed in a very short and broad cylinder, the top and bottom of which were both formed of disks rotating at a certain angular velocity surrounded by an annulus rotating at a different rate of rotation. In \S4.3 of Ref.~\cite{Rabaud} they considered a linear model problem in which a planar shear zone was formed in the center region of a rectangular duct of which the two halves moved in opposite directions. 

In this paper another type of laminar shear layer will be considered, namely the plane shear layer which forms in the interaction zone between two co-current and fully-developed plane Couette flows\footnote{The flow beneath a flat-bottomed ship, e.g. a tanker or bulk carrier, which travels with a small underkeel clearance to a flat sea floor can be described as a turbulent Couette flow at full scale but may be laminar at model scale in a towing tank\cite{Gourlay}. Two co-current Couette flows, as considered here, therefore mimic the interaction of the flows beneath two adjacent ships in side-by-side operation.}.
Contrary to the classical free shear layer \cite{Lock,Panton,White}, the width of the new shear layer is independent of the streamwise position. As we shall see, however, the extent of the shear zone varies in the direction normal to the streams and thus shares some features with the linear model problem solved earlier by Rabaud and Couder \cite{Rabaud}. In the present study we furthermore aim to compare the analytically derived solution with a numerical solution of the full three-dimensional Navier-Stokes equations in order to justify the inherent assumption of unidirectional flow.

\section{Mathematical formulation}

Let us consider the shear layer which develops between two streams with velocities $U_2$ and 
$U_1$ in the $(x,y)-$plane. While $U_2$ and $U_1$ are taken as constants in the classical shear layer theory, let us now assume that both $U_2$ and $U_1$ vary linearly with the lateral position.

We will assume that the flow field is steady and fully developed in the streamwise direction, 
i.e. the three velocity components $U$, $V$, and $W$ are functions only of the two coordinates 
$y$ and $z$. The Navier-Stokes equations therefore simplify to:

\begin{subequations}
\begin{eqnarray}
V \py U + W \pz U &=& \nu \left( \py^2 U + \pz^2 U\right) ;
\label{umomen} \\
V \py V + W \pz V &=& -\rho^{-1} \py P + \nu (\py^2 V + \pz^2 V) ;
\label{vmomen} \\
V \py W + W \pz W &=& -\rho^{-1} \pz P + \nu (\py^2 W + \pz^2 W) ;
\label{wmomen} \\
\py V + \pz W &=& 0 .
\label{cont}
\end{eqnarray}
\end{subequations}
From the above system of partial differential equations, it is observed that the `secondary motion', 
i.e.\ the flow $(V, W)$ in the cross-sectional plane, is independent of the streamwise velocity component $U$. 
This flow is governed by eqs.\ (\ref{vmomen}) $-$ (\ref{cont}) for which the zero solution $V = W = P = 0$ is a valid 
and consistent solution.
If so, equation (\ref{umomen}) 
for the $x-$component of the velocity vector simplifies to the 
Laplace equation:

\begin{equation}
\py^2 U + \pz^2 U = 0 .
\label{laplace}
\end{equation}

A flow like this can be realized in the interaction zone between two co-current plane Couette flows, 
as shown in figure \ref{domain}(b). The lower wall at $z = 0$ is fixed whereas the upper wall at $z = 2h$ 
moves in the positive $x-$direction with a speed which is discontinuous at $y = 0$ such that the wall velocity 
is $U_-$ and $U_+$ for $y < 0$ and $y > 0$, respectively. Explicitly, the boundary conditions are 
\begin{align*}
  U(x, y, z=0)=&0, \\
  U(x, y>0, z=2h)=&U_+, \\
  U(x, y<0, z=2h)=&U_-.
\end{align*}
The velocities far away from the shear zone will therefore tend to the linear Couette flow profiles $U_1 = z U_- /2h$ and $U_2 = z U_+ /2h$ when we assume the flow to be steady and fully developed.

\section{Numerical simulation}

Let us first solve the full Navier-Stokes equations in three-dimensional space and time for a flow 
configuration as shown in figure \ref{domain}(b). The Reynolds number $Re_- = U_- \ h/2\nu$ based on 
half of the speed $U_-$ of the fastest moving wall and half of the wall distance $2h$ was taken as 260 
and the wall-speed ratio was $U_- \ /U_+ = 2.0$. This Reynolds number is well below the subcritical 
transition Reynolds number below which the plane Couette flow is known to be stable.

The Navier-Stokes equations for an incompressible and isothermal flow are solved
using a parallel Finite Volume code called MGLET \cite{Manhart}. The code uses staggered Cartesian 
grid arrangements. Spatial discretization of the convective and diffusive fluxes
are carried out using a $2^\textrm{nd}$-order central-differencing scheme. The
momentum equations are advanced in time by a fractional time stepping using a
$2^\textrm{nd}$-order explicit Adams-Bashforth scheme. The Poisson equation for the
pressure is solved by a full multi-grid method based on pointwise
velocity-pressure iterations. The computational grid is divided into an
arbitrary number of subgrids that are treated as dependent grid blocks in
parallel processing. In the present study, the size of the computational domain $L_x \times L_y \times L_z$
and the number of grid points in each coordinate direction are $50.24h \times 16.8h \times 2h$ 
and $256 \times 256 \times 64$, i.e. comparable to the \textit{turbulent} plane Couette flow 
simulation of Bech et al.~\cite{Bech}.
Uniform grid spacing is adopted in the streamwise and the 
spanwise directions, while a non-uniform mesh is used in the wall-normal direction. Periodic boundary
conditions are employed in the streamwise and spanwise directions. No-slip and
impermeability conditions are imposed on the walls.

\begin{figure}
\centering
\includegraphics[width=0.45\textwidth]{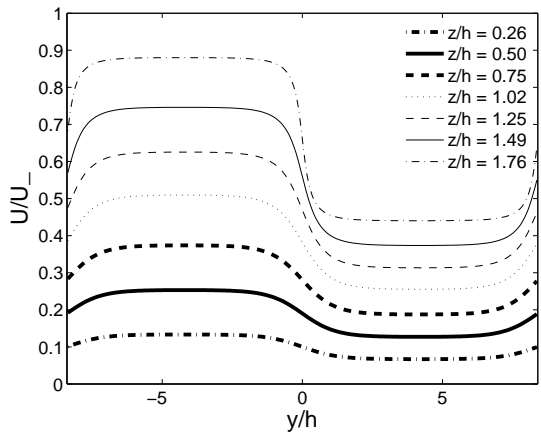}
\caption{Spanwise variations of the streamwise velocity $U/U_-$ at different wall-normal positions $z/h$.
Numerical solution of the full Navier-Stokes equation.}
\label{mixing}
\end{figure}

\begin{figure}
\includegraphics[width=.5\textwidth]{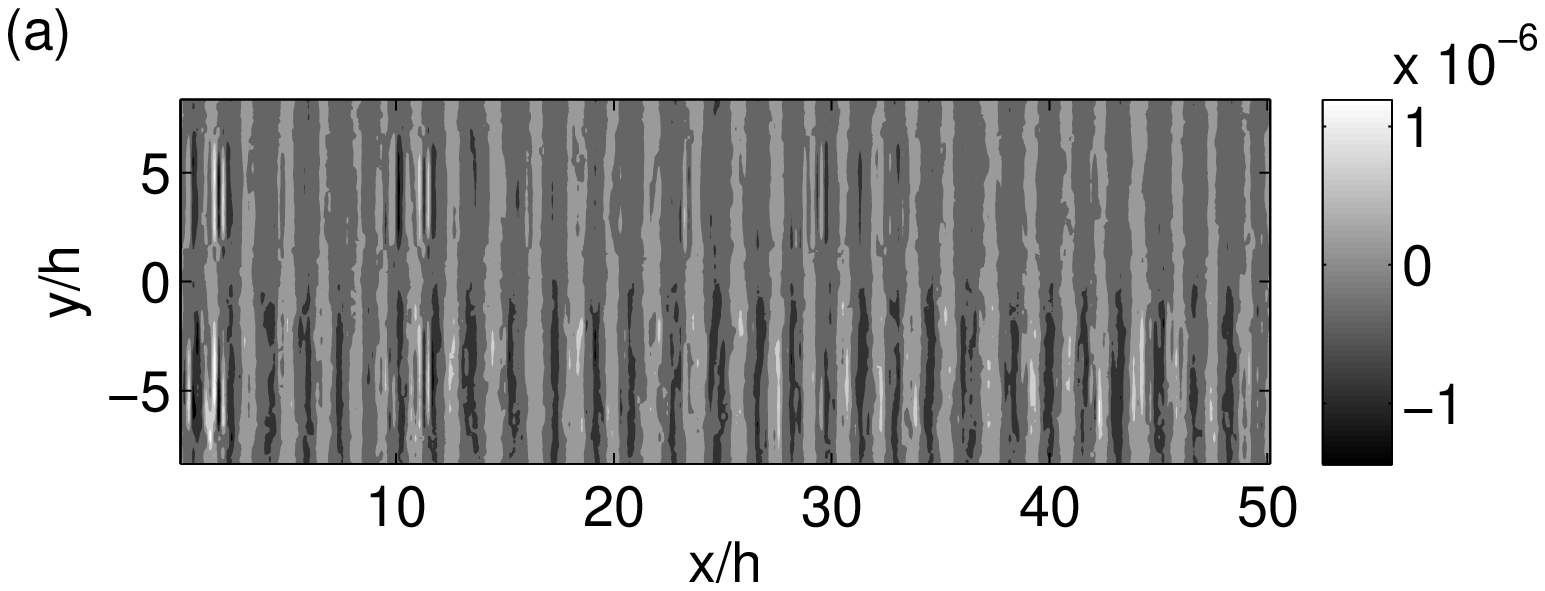}\\
\includegraphics[width=.5\textwidth]{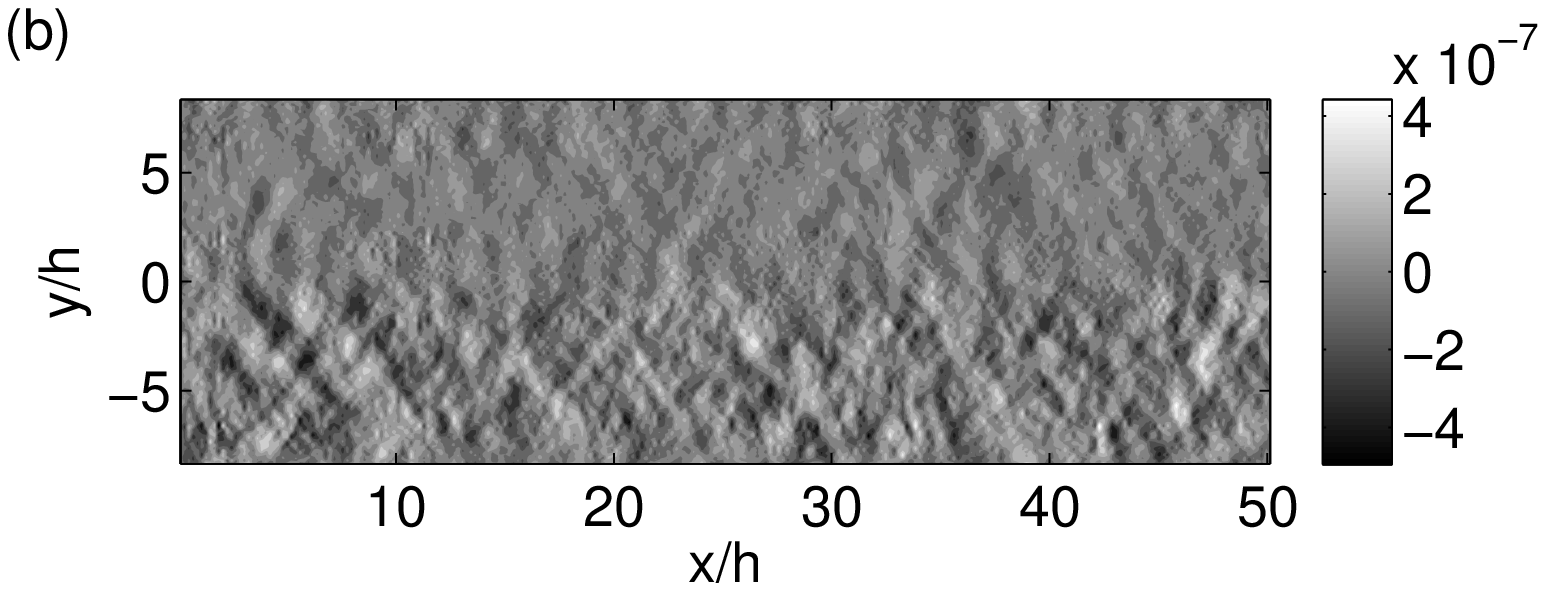}
\caption{Normalized (a) wall-normal velocity $W/U_-$ and (b) spanwise velocity $V/U_-$ contours
in the mid-plane. Numerical solution of the full Navier-Stokes equation.}
\label{vw}
\end{figure}

The numerical solution of the unsteady Navier-Stokes equations converged to a steady state.
The present computations thus yielded a flow field which is steady
and fully developed in the streamwise direction. This steady flow field is illustrated in 
figure \ref{mixing}, where the streamwise velocity profiles are plotted along the span.
It can be observed that in regions away from the shear zone the streamwise velocity varies
monotonically in the wall-normal direction in a linear fashion similar to laminar plane Couette flows.
The shear layer width is minimal just below the moving plates where the velocity gradient 
($\partial U/\partial y$) is high. Due to viscous diffusion the steepness of the velocity profiles 
is gradually reduced towards the stationary wall.
In figure \ref{vw} non-dimensionalized wall-normal velocity $W$ and spanwise velocity $V$
at the channel mid-plane are shown. These results clearly illustrate that these two
velocity components are practically zero in magnitude and hence negligible. 

\section{Analytical solutions}

The results from the numerical solution of the full Navier-Stokes equation showed beyond any doubt 
that the secondary velocity components $V$ and $W$ are negligible for all practical purposes. The 
streamwise velocity component $U$ is therefore governed by the Laplace equation (\ref{laplace}) for 
which an analytical solution now will be sought.
Assuming separation of variables $U(y,z)=Y(y)Z(z)$ we obtain the decoupled equations
\be
  Y'' = -k^2 Y; ~~ Z'' = k^2 Z
\ee
where $k$ is a constant. The general solution to the equation for $Y$ in terms of $\cos(ky)$ and $\sin(ky)$ and periodic boundary conditions $Y(y+2L)=Y(y)$ imply a discretization of $k$ according to $k = n\pi/L$ where $n\in \mathbb{N}$. 
Standard theory of Fourier series to satisfy all boundary conditions gives the answer
\begin{align}
  \frac{U^{\rm pbc}(y,z)}{\langle U \rangle} = \frac{z}{2h} + \frac{2\Lambda}{\pi} 
  \sum_{n=0}^\infty\frac{\sinh\frac{(2n+1)\pi z}{L}}{\sinh\frac{(2n+1)2\pi h}{L}}\frac{\sin\frac{(2n+1)\pi y}{L}}{2n+1}\label{periodicFourier}
\end{align}
where the average of the two wall velocities is $\langle U \rangle = {\textstyle \frac1{2}}(U_-+U_+)$. One notes that the 
velocity, when reduced by the average of the two wall velocities, depends on $U_-$ and $U_+$ exclusively through a single parameter 
\be
  \Lambda = (U_+-U_-)/\langle U \rangle
\ee
according to (here and henceforth $\langle U \rangle \neq 0$)
\be\label{scaling}
  \frac{U(y,z)}{\langle U \rangle} = \frac{z}{2h} + \Lambda f_{h,L}(y,z).
\ee
The first term is obviously the simple linear Couette flow profile for a single wall moving at velocity $\langle U\rangle$. 
All information about the shear layer is now contained in the function $f_{h,L}(y,z)$. We will call the second term 
the `asymmetry term' of $U(y,z)$.

\begin{figure}[tb]
  \begin{center}
  \includegraphics[width=\columnwidth]{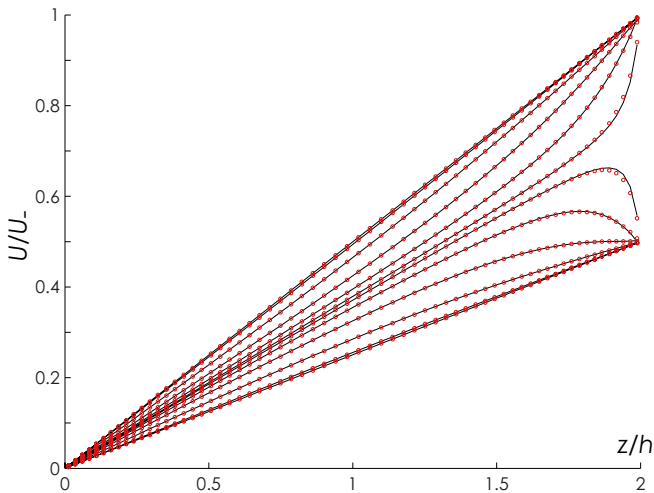}
  \caption{(Color online) Streamwise velocity $U(y,z)/U_-$ variations along the wall-normal direction $z/h$. Full numerical solution (circles) compared with analytical Eq.~(\ref{finalresult}) (solid lines) at spanwise positions (top to bottom) $y/h= -4.2$, $-2.1$, $-0.95$, $-0.42$, $-0.16$, $-0.03$, $0.03$, $0.16$, $0.42$, $0.95$, $2.1$, $4.2$. The top two and bottom two lines nearly coincide.}
  \label{fig_numAnaComp}
  \end{center}
\end{figure}

From Eq.~(\ref{periodicFourier}) we may obtain the solution for the case of Neumann boundary conditions at $y=\pm \infty$ by taking the limit $L\to \infty$. The sum then becomes an integral according to the Euler-Maclaurin formula (e.g.~Ref~\cite{Abramowitz64}).
It is possible to solve 
the resulting
integral in closed form, 
however, the Neumann solution is more elegantly obtained using complex analysis. 

The boundary value problem is conformally invariant under the Schwarz-Christoffel-type map
\be
  \eta = y+iz = \frac{4h}{\pi}\mathrm{artanh}\zeta; ~~  \zeta = \tanh \frac{\pi \eta}{4h}.
\ee
The boundaries of the original problem are now mapped onto the real $\zeta$ axis so that $z=0$ maps to $\zeta\in (-1,1)$, the boundary $z=2h$, $y>0$ where the velocity is $U_+$ is mapped to $\zeta \in (1,\infty)$ and $z=2h$, $y<0$ where the velocity is $U_-$ to $\zeta \in (-1,-\infty)$. All other points in the original strip domain are mapped conformally to the upper half of the complex $\eta$ plane. Seeking $w(\eta)$ so that $U(\eta)=\mathrm{Im}[w(\eta)]$, we can instead find a function $w(\zeta(\eta))$ satisfying the boundary conditions. Such is the case for 
\be
  w(\zeta) = \frac{U_-}{\pi}\log(\zeta+1)-\frac{U_+}{\pi}\log(1-\zeta)
\ee
and with some manipulation we find
\be\label{finalresult}
  \frac{U^\text{Nbc}(y,z)}{\langle U \rangle} = \frac{z}{2h} + \frac{\Lambda}{\pi}\arctan\left[\tanh\left(\frac{\pi y}{4h}\right) \tan \left(\frac{\pi z}{4h}\right)\right].
\ee
It is easy to see that simple linear Couette profiles are obtained far aside of the shear layer as they should, noting that for $y\gg h$, $\tanh(\frac{\pi y}{4h}) \to \mathrm{Sg}(y)$, i.e., the signum function.

\section{Discussion}

The closed-form analytical solution (\ref{finalresult}) is applicable when the flow is subjected to Neumann boundary 
conditions in the spanwise direction, whereas the solution (\ref{periodicFourier}) applies for spanwise periodicity. 
We first verified that the latter series solution with $2L = 16.8h$ coincided with the closed-form solution 
(\ref{finalresult}) in the shear zone. A comparison between the latter and the full Navier-Stokes solution is provided 
in figure \ref{fig_numAnaComp}. Recall that the width of the computational domain on which the Navier-Stokes equations 
were integrated was $L_y = 16.8h$ whereas the displayed velocity profiles only span the central $8.4h$ of the shear zone. 
The numerical and analytical solutions agree to a relative error of order $0.01\%$, except in the immediate vicinity of the discontinuity in the wall velocity where deviations of about $1\%$ are found.
Hence the shear layer is excellently described by the analytical expression given as the second term on the right hand side of Eq.~(\ref{finalresult}), and an analytical expression for its width variations in the wall normal direction can be found as follows. 

\begin{figure}[tb]
  \begin{center}
  \includegraphics[width=.4\textwidth]{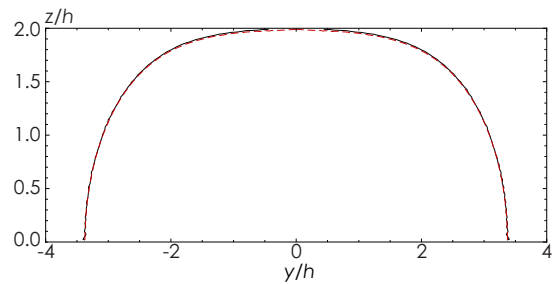}
  \caption{(Color online) Numerical shear layer thickness for $\delta=0.01$ (solid line) compared with approximation Eq.~\eqref{mixingthickness} (dashed line).}
  \label{fig_thickness}
  \end{center}
\end{figure}

Let us define the shear layer symmetrically about $y=0$, that is, it is the area within which the asymmetry term of 
$U(y,z)$ in Eq.~(\ref{scaling}) differs from its asymptotic form far from the shear region by more than a small relative measure $\delta$, typically $1\%$. Far from the shear layer the asymmetry term tends to that for simple Couette flow, $\mathrm{Sg}(y)\Lambda z/4h$, so according to equation (\ref{finalresult}) the edges of this shear layer are given as the positive and negative $y$ solution of
\[
  \frac{\Lambda}{\pi}\arctan\left[\tanh\left(\frac{\pi y}{4h}\right) \tan \left(\frac{\pi z}{4h}\right)\right]=\mathrm{Sg}(y)(1-\delta)\frac{\Lambda z}{4h}.
\]
It is clear that in the laminar regime the thickness of the shear layer is completely independent of the velocities 
$U_-$ and $U_+$. The shear layer is quite obviously symmetric around $y=0$ such as we have defined it. We take the tangent of either side and expand to linear order in $\delta$. Noting that for $|\tanh x|$ close to unity we may use the approximation, $\tanh x\approx \mathrm{Sg}(x)[1-2\exp(-2x)]$, we find the approximate shear layer thickness $d_\text{layer}$ as a function of $z$ as
\be\label{mixingthickness}
  \frac{d_\text{layer}}{h} \approx -\frac{2}{\pi}\log\left(\frac{\pi z}{4h}\frac{\delta}{\sin\frac{\pi z}{2h}}\right).
\ee
For small $\delta$ this is an excellent approximation except 
in the immediate vicinity of the discontinuity
as shown in figure \ref{fig_thickness}.

According to its definition, the shear layer thickness vanishes at $z=2h$, but not near the stationary wall at $z=0$. This is a consequence of the definition of the shear layer, which is the region within which the relative difference between the real velocity profile and the asymptotic (plane Couette) flow at infinity is above some threshold. Thus, although all absolute velocities tend to zero at $z=0$, the values of $U$ taken at different $y$ relative to each other remain finite and nonzero in this limit, hence the shape of the shear layer as shown in Fig.~\ref{fig_thickness}.

\begin{figure}[tb]
  \begin{center}
  \includegraphics[width=.5\textwidth]{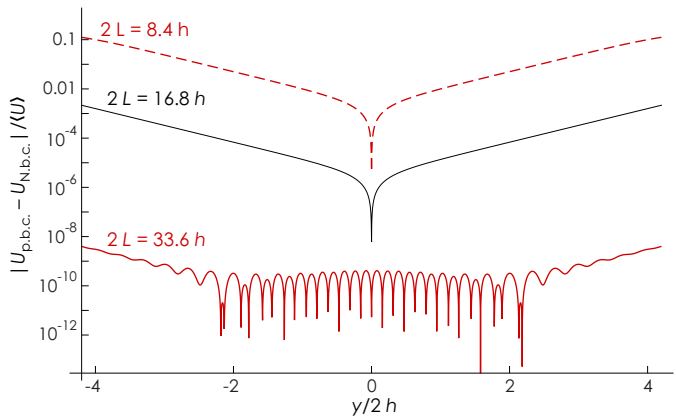}
  \caption{(Color online) Absolute difference between $U$ in the cases of periodic (period $2L$) and Neumann ($L=\infty$) boundary conditions, whose analytical expressions are Eqs~\eqref{periodicFourier} and \eqref{finalresult}, respectively. This quantifies the interaction of shear layers at $y=0$ and $y=\pm L$. We compare the period used in this paper, $2L=16.8h$, with half and twice this period. For the longest period the difference is at the level of the cutoff error from including just $80$ terms in the Fourier series, Eq.~\eqref{periodicFourier}, giving rise to the oscillating behavior in this case}.  
  \label{fig_difference}
  \end{center}
\end{figure}

With our ratio $L/2h = 8.4$ there is very little mutual influence between the shear profile at $y=0$ and that at $y=\pm L$ appearing due to the periodic boundary conditions. To quantify we plot the difference between the periodic and Neumann boundary condition velocity profiles, Eqs.~\eqref{periodicFourier} and \eqref{finalresult}, respectively, in Fig.~\ref{fig_difference}. For comparison the same is plotted when the period $L$ is halved and doubled. For our choice of $L/h$ ratio the influence of the periodic boundary conditions remains below the $1\%$ level all the way out to the half-way points $y=\pm L/2$. Doubling the spanwise period further reduces the influence by five orders of magnitude.

\section{Concluding remarks}

We have considered the shear layer between two 
co-current streams with constant vorticities $a_2 = U_+ \ /2h$ and 
$a_1 = U_- \ /2h$, respectively. The numerical solution of the full 3D Navier-Stokes equation first showed that 
$V$ and $W$ were totally negligible away from the plane of the velocity discontinuity. The streamwise momentum equation therefore simplified to a Poisson equation 
(\ref{laplace}) for $U(y,z)$. By assuming spanwise periodicity, the analytical solution (\ref{periodicFourier}) was 
derived. As the period tends to infinity, the analytic solution simplified to the closed-form solution (\ref{finalresult}). 

The width of the shear zone increased monotonically from the moving split wall towards the fixed bottom plane where the width $2d_\text{layer} \approx 6.8h$. Outside of the shear region, the two co-current streams are constant-vorticity 
Couette flows where the spanwise vorticity $\omega_y = \P U/ \P z$ is generated by the wall motion and diffused downwards. In the shear zone, $\omega_y$ of the fastest-moving fluid is reduced in the lower part and increased as the moving wall is approached, whereas $\omega_y$ of the low-speed flow is enhanced in the lower part but decreases near the upper wall. In addition, the shear zone gives rise to a wall-normal vorticity $\omega_z = - \P U/ \P y > 0$ and the highest level of $\omega_z$ is reached in the vicinity of the moving wall at $y/h = 0$. In the present flow the variations of $\omega_y$ and $\omega_z$ are governed solely by viscous diffusion. It is noteworthy that in order to maintain $\omega_x = 0$, tilting of $\omega_y$ and $\omega_z$ by means of velocity strains, i.e. $\omega_y \ \P U/ \P y$ and $\omega_z \ \P U/ \P z$, respectively, are exactly outweighed.

In order to check the stability of the present flow, the numerical integration of the full Navier-Stokes equation 
was repeated with random noise superimposed on the initial flow field. The solution eventually evolved to the same 
steady state as before, thereby suggesting that the laminar flow is stable at the Reynolds number $Re_- = 260$ 
considered here.

\begin{figure}[b!]
\centering
\includegraphics[width=3.0in]{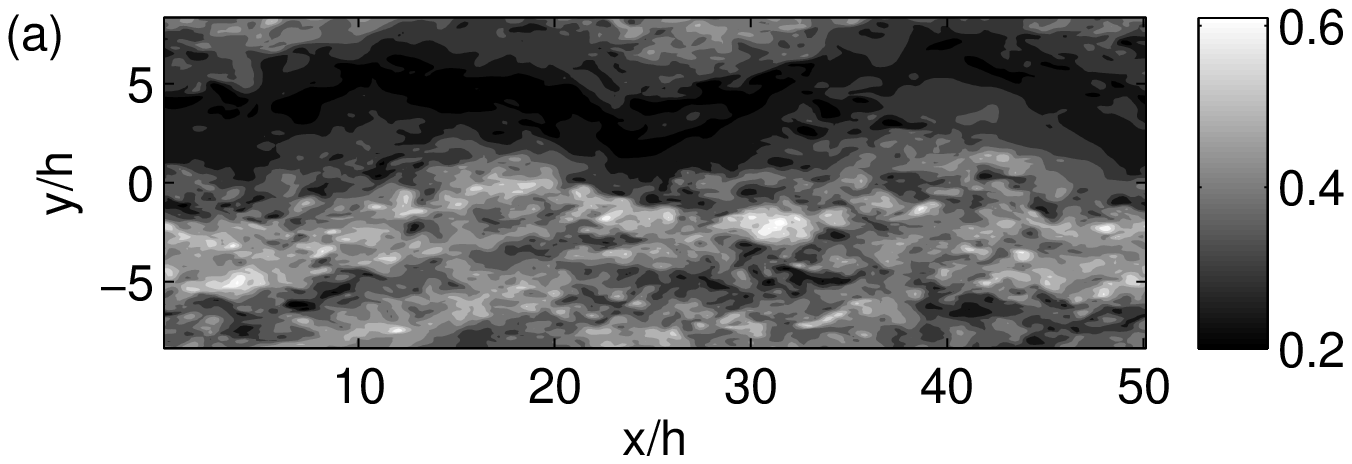}\\
\includegraphics[width=3.0in]{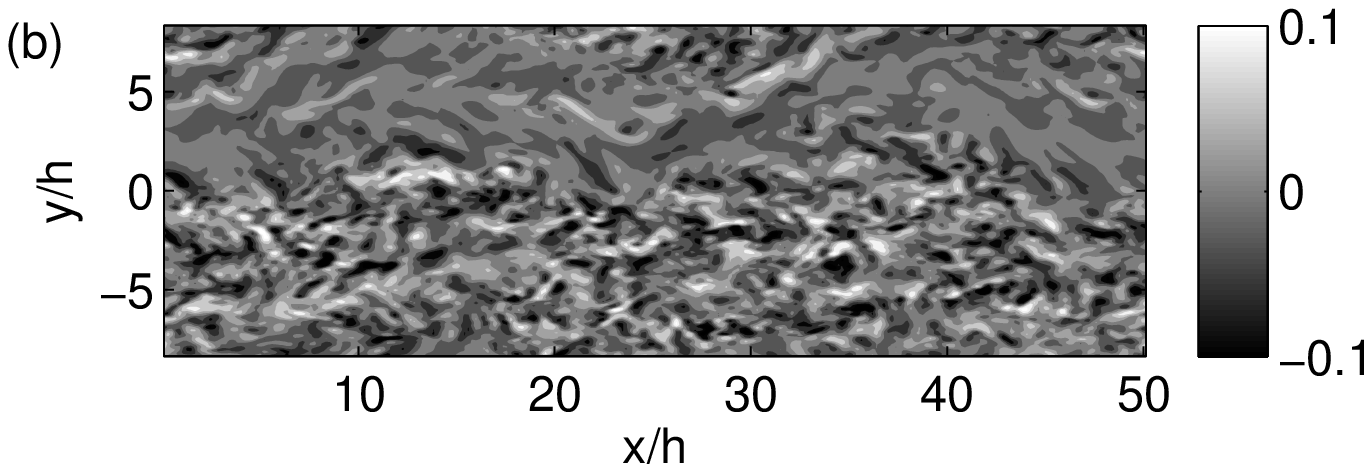}\\
\includegraphics[width=3.0in]{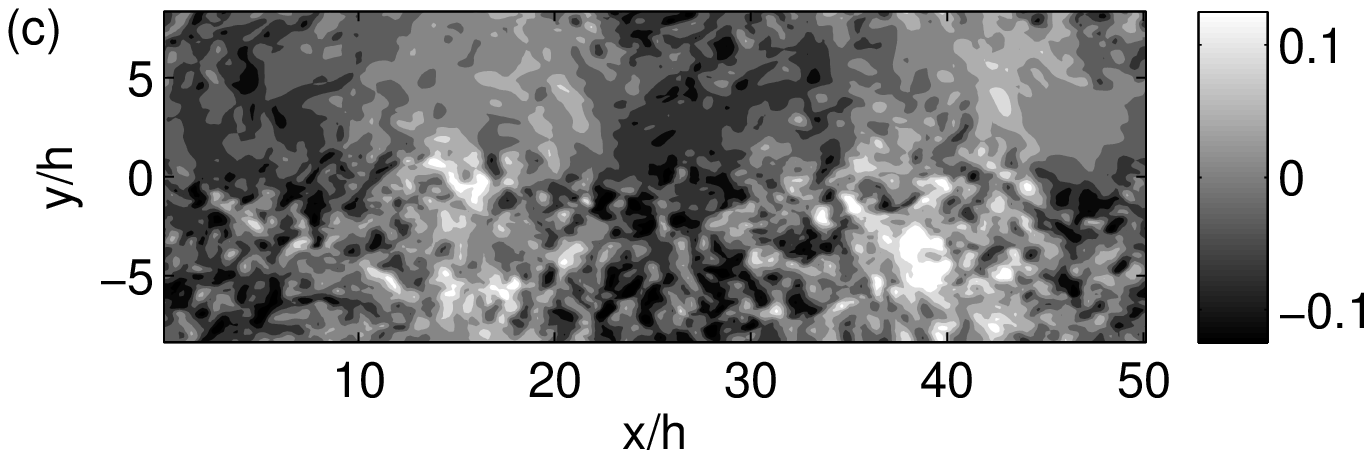}
\caption{Instantaneous velocity components from the DNS data of Narasimhamurthy et al.\ \cite{ETC13} where $U_- \ /U_+ = 5.0$ and $Re_- = 1300$ and $Re_+ = 260$. All profiles are extracted from the mid-plane. (a) streamwise velocity $U/U_-$; (b) wall-normal velocity $W/U_-$; (c) spanwise velocity $V/U_-$.}
\label{uvw}
\end{figure}

The stability of the shear-layer at a higher Reynolds number was further investigated by Narasimhamurthy et al. \cite{ETC13}, where $Re_-$ was increased to 1300 while keeping the $Re_+$ equal to $260$. The computational domain and the number of grid points were the same as before. This resulted in a velocity ratio of $U_-/U_+ = 5.0$. The higher $Re_-$ considered in \cite{ETC13} is well above the subcritical transition Reynolds number of $300$--$370$ (see Schneider et al. \cite{Schneider} and Tuckerman \& Barkley \cite{Tuckerman}) for which the plane Couette flow is fully turbulent. Thereby, the interface between a laminar and a turbulent plane Couette flow was studied in Narasimhamurthy et al.\ \cite{ETC13} rather than the present shear layer between two laminar Couette flows. The higher $Re_-$, i.e., turbulence, had a distinct effect on the interface where shear-layer instabilities were developed resulting in large-scale interactions between the turbulent and the nominally non-turbulent part flow. Such a large-scale mixing-zone is shown in figure \ref{uvw} where instantaneous velocity components from \cite{ETC13} are plotted. The secondary-flow in figure \ref{uvw}(c) together with the meandering-motion or the unsteadiness in the other two velocity components as depicted in figure \ref{uvw}(a, b) clearly indicate that the shear-layer is no longer stable under turbulent conditions. Thereby, a mixing-layer was established in \cite{ETC13} rather than the present stable shear-layer (see \cite{ETC13} for more details on the momentum transfer and turbulent diffusion mechanisms occurring in the mixing-zone).

The dynamics of a laminar-turbulent interface has also been   
investigated in a plane Couette flow configuration at Reynolds   
numbers in the range from 180 to 650 by Duguet et al.~\cite{duguet11}. Contrary  
to the co-current laminar-turbulent Couette flow \cite{ETC13}, initial   
perturbations were introduced into a conventional plane Couette flow 
driven by continuous wall motion.  Sufficiently strong perturbations 
at sufficiently high Reynolds numbers led to turbulence localized in 
the spanwise direction which enabled detailed explorations of the 
interface dynamics.

The Taylor-Couette flow between rotating cylinders approaches the plane Couette flow in the limit of large radii and small gaps, as shown for instance by Faisst and Eckhardt \cite{faisst00}. The present flow configuration can thus be considered as a limiting case of a Taylor-Couette flow with different rotational speeds of the upper and lower parts of the driving cylinder. 

Another analogy to the currently considered geometry is that considered experimentally, e.g., by Burin and co-workers, in which the end-walls of a Taylor-Couette set-up consist of two independently moving rings rotating with the inner and outer cylinders, respectively \cite{burin06}. Again the present flow configuration is obtained as a limiting case.

The shear layer which forms between two co-current plane Couette flows shares some similarities with the shear zone formed in the junction between the two halves of a fluid-filled duct which 
move in opposite directions. Rabaud and Couder \cite{Rabaud} derived an analytic solution for the latter problem by assuming that the secondary motion was negligible and thereafter solving Laplace's equation (\ref{laplace}). The width of the resulting shear zone was largest midway between the two parallel duct walls, whereas the width of the present shear layer increased all the way from the splitted moving plate to the stationary wall. Another distinguishing feature of the present flow is that linear Couette flow profiles are recovered outside of the shear zone such that both spanwise vorticity and viscous shear still exist. Outside of the shear zone in the flow analysed by Rabaud and Couder \cite{Rabaud}, the fluid was conveyed as a solid body along with the moving halves of the duct and neither vorticity nor viscous shear stresses prevailed.  

A new shear layer has been introduced in this paper. Contrary to the classical shear layer, the width of the shear zone varies in the direction perpendicular to the shear but is independent of the streamwise direction. An analytical solution has been derived which compared perfectly well with accurate numerical solutions of the three-dimensional Navier-Stokes equations. The solution turned out to be independent of the fluid viscosity, which implies that the solution is valid for all Reynolds numbers sufficiently low for the flow to remain stable.

\subsection*{Acknowledgements}

This work has received support from The Research Council of Norway 
(Programme for Supercomputing) through a grant of computing time.
We thank anonymous referees for helpful suggestions.

\end{document}